\documentclass[%
 reprint,
superscriptaddress,
 amsmath,amssymb,
 aps,
]{revtex4-1}

\usepackage{graphicx}
\usepackage{dcolumn}
\usepackage{bm}
\usepackage{hyperref}

\begin{document}


\title{Ray mapping approach for the efficient design of continuous freeform surfaces}

\author{Christoph B\"osel}
 \email{christoph.boesel@uni-jena.de}
\author{Herbert Gross}%
\affiliation{Friedrich-Schiller-Universit\"at Jena, Institute of Applied Physics, Abbe Center of Photonics, 07743 Jena, Germany}%

\begin{abstract}
The efficient design of continuous freeform surfaces, which transform a given source into an arbitrary target intensity, remains a challenging problem. A popular approach are ray-mapping methods, where first a ray mapping between the intensities is calculated and in a subsequent step the surface is constructed. The challenging part hereby is the to find an \textit{integrable} mapping ensuring a \textit{continuous} surface. Based on the law of reflection/refraction and the well-known integrability condition, we derive a general condition for the surface and ray mapping for a collimated input beam. It is shown that in a small-angle approximation a proper mapping can be calculated via optimal mass transport. We show that the surface can be constructed by solving a linear advection equation with appropriate boundary conditions. The results imply that the optimal mass transport mapping is approximately integrable over a wide range of distances between the freeform and the target plane and offer an efficient way to construct the surface by solving standard integrals. The efficiency is demonstrated by applying it to two challenging design examples.
\begin{description}
\item[PACS numbers]
 \verb+42.15.-i, 42.15.Eq+.
\end{description}
\end{abstract}

\maketitle


\section{Introduction}
\label{sec:Introduction}

In recent years much progress has been made in the field of freeform surface design without the assumption of symmetries \cite{ries02, Wu13_1, Wu13_2, Wu13_3, Wu14_1, Wu14_2, Prins14_1, Brix15_1, Oliker02_1, Fou10_1, Can13_1, Ma15_1, Mic11_1, Oli13_1, Oli14_1, Oli14_2,Rub07_1, Bau12_1,Brun13_1, Feng13_1, Feng13_2}. The goal of these design methods is the solution of the so called inverse problem of nonimaging optics. This means that for given \textit{arbitrary} source and target intensities $I_S (x,y)$ and $I_T (x,y)$ one or more freeform surfaces have to be calculated, which map the intensities onto each other. Especially the design of \textit{continuous} freeform surfaces, on which we will concentrate on in the following, is a challenging problem and of great interest for practical applications.\\
The first successful method at calculating a continuous freeform surface utilizing a complex target intensity was developed by Ries and Muschaweck \cite{ries02}, but unfortunately the numerical method was not published \cite{ries02}. Their approach is able to handle the far field design problem for single freeform surfaces illuminated by a point source \cite{ries02}.\\
Nowadays, many other methods have been developed by different research groups. A quite popular approach are the so-called Monge-Ampère methods \cite{Wu13_1, Wu13_2, Wu13_3, Wu14_1, Wu14_2, Prins14_1, Brix15_1}. They are based on the modelling of the design problem by a nonlinear partial differential equation of Monge-Ampère type and solving it with sophisticated numerical techniques. These methods are able to handle the design problem in the far field for intensity control of point sources \cite{Wu13_1, Wu14_2, Brix15_1} and collimated beams \cite{Wu13_3, Wu14_2, Prins14_1} as well as intensity and phase control with double freeform surfaces \cite{Wu14_1}.\\
Another popular approach for the single freeform surface design with point sources is the supporting ellipsoids method developed by Oliker \cite{Oliker02_1}. With this method a freeform mirror is constructed by putting a point source in the focal point of a unification of ellipoids, whereby every ellipsoid has a different position of the second focal point on the target plane to build the required intensity pattern. The challenge of this method is the calculation of a smooth freeform surface by the unification of the ellipsoids. Therefore it was further developed by other research groups \cite{Fou10_1,Can13_1} and generalized to calculate freeform lenses \cite{Mic11_1}. It can handle the far field as well as the near field design problem.\\
Also quite often used are ray-mapping techniques \cite{Fou10_1, Ma15_1, Rub07_1, Oli13_1, Oli14_1, Oli14_2, Bau12_1, Brun13_1, Feng13_1, Feng13_2}, which are frequently based on the calculation of a ray mapping between the source and the target intensity and a subsequent construction of the freeform surface \cite{Fou10_1, Ma15_1, Rub07_1, Oli13_1, Bau12_1,Brun13_1, Feng13_1, Feng13_2}. The aim and challenging part of these methods is to find an \textit{integrable} ray mapping, which allows the calculation of a \textit{continuous} surface.\\
The approach we will concentrate on is a subgroup of the ray-mapping techniques, which are called optimal mass transport (OMT) methods \cite{Bau12_1,Brun13_1, Feng13_1, Feng13_2}. These gained some interest in recent years and are partly based on the mathematical concept of optimal mass transport as explained in the following paragraph. They can handle the design problem of a single and double freeform surfaces for intensity control \cite{Bau12_1, Brun13_1} as well as double freeform surfaces for intensity and phase control \cite{Feng13_1, Feng13_2}.
The connection between mass transport and freeform design was also discussed in \cite{Oli11_1}.\\
This approach for the freeform surface design consists of two separate steps. In the first step a ray mapping between the given input and output intensities is calculated via OMT. In the second step the freeform surface is constructed with the help of the law of refraction/reflection and the well-known integrability condition, which ensures the continuity of the surface.\\
The difference between the OMT methods mentioned above is the second step. In first approach by Bäuerle \textit{et al.} the freeform is constructed by an optimization procedure \cite{Bau12_1, Brun13_1}, while the second approach by Feng \textit{et al.} uses a simultaneous point-by-point construction method to design a double freeform surface \cite{Feng13_1, Feng13_2}.\\
Since these attempts seem to be quite successful but do not give theoretical insights about the integrability of the OMT map, we want to clarify this point in our work for a single freeform illuminated by a collimated beam. This will be done by deriving a condition for an integrable map and showing that it can be fulfilled (approximately) by the OMT map. Based on our findings, we present an efficent and easy-to-implement numerical freeform surface construction technique differing from the previously published OMT methods \cite{Bau12_1,Brun13_1, Feng13_1, Feng13_2}. To do so this paper is structured as follows.\\
In section \ref{sec:Design method} after a short introduction to the OMT and a presentation of its basic properties, we will derive from the law of reflection/refraction and the integrability condition a \textit{general} condition for an integrable ray mapping and its corresponding surface for collimated input beams. It will be shown that
in a small-angle approximation this condition can be fulfilled by using an OMT mapping and therefore the freeform surface design process indeed decouples into the two steps described above. Thereby it will be shown that the freeform surface can be constructed from a linear advection equation with appropriate boundary conditions. In section \ref{sec:Numerical Algorithm} we then argue that the advection equation for the freeform construction can be solved by simple integrations, which is different to the OMT freeform design methods mentioned above and implies the approximate integrability over a wide range of freeform-target plane distances. The efficiency of this approach is then demonstrated in section \ref{sec:Examples} by applying it to two challenging design examples, followed by a discussion of our results in section \ref{sec:Conclusion}.

\section{Design method}
\label{sec:Design method}

\subsection{Optimal mass transport}
\label{sec:Optimal mass transport}

\begin{figure*}
\includegraphics[width=\textwidth]{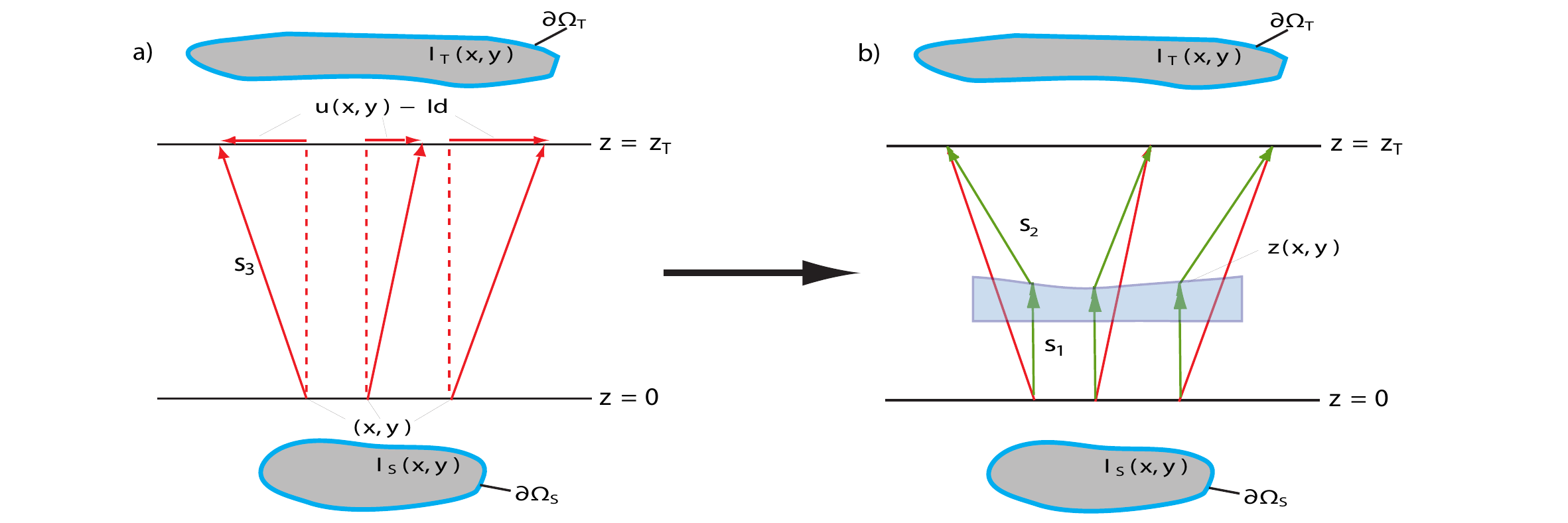} 
  \caption{a) For the given input intensity $I_S (x,y)$ and output intensity $I_T (x,y)$ the OMT ray mapping $\mathbf{u}(x,y)$ is calculated. The mapping defines a vector field $\mathbf{s}_3$ between the source plane $z=0$ and the target plane $z=z_T$. $\partial \Omega_S$ and $\partial \Omega_T$ are the source and target intensity boundaries, respectively.
b) In the second process step the freeform surface $z(x,y)$ is constructed in a way that it is continuous and redirects the incoming collimated beam, described by the vector field $\mathbf{s_1 }$, according to the given OMT map. Because of energy conservation, the boundary of the freeform corresponds to the shape of the source intensity $\partial \Omega_S$.}
\label{fig:Geometrie}
\end{figure*}

The problem statement of OMT, also called the Monge-Kantorovich problem, is as follows: two positive density functions $I_S (\mathbf{x})$ and $I_T (\mathbf{x})$  with
\begin{equation}\label{eq:1}
\int_{\mathbb{R}_2}I_S (\mathbf{x}) d\mathbf{x}= \int_{\mathbb{R}_2}I_T (\mathbf{x}) d\mathbf{x}
\end{equation}
have to be mapped onto each other according to the Jacobi equation
\begin{equation}\label{eq:2}
det(\nabla \mathbf{u}(\mathbf{x}))I_T (\mathbf{u}(\mathbf{x}))=I_S (\mathbf{x})
\end{equation}
with a smooth, bijective mapping $\mathbf{u}(\mathbf{x})$. If $M$ is defined as the set of mappings fulfilling equation (\ref{eq:2}), we are searching for a mapping minimizing the transport cost according to the Kantorovich-Wasserstein distance
\begin{equation}\label{eq:3}
d(I_S, I_T)^2 = inf_{\mathbf{u}\in M} \int |\mathbf{u}(\mathbf{x})-\mathbf{x}|^2 I_S (\mathbf{x}) d\mathbf{x},
\end{equation}
whereby $inf_{\mathbf{u}\in M}$ denotes the mapping for which the integral has its minimal value. This mapping, which is defined by (\ref{eq:1}), (\ref{eq:2}) and (\ref{eq:3}), has the useful property that it is unique \cite{Bre02_1} and it is characterized by a vanishing curl \cite{Hak04_1}. The latter property will be important for our findings in the next subsection.\\
In the special case of freeform surface design considered here, the densities $I_S (\mathbf{x})$ and $I_T (\mathbf{x})$ correspond to the source and target intensities with the units $W \cdot m^{-2}$ (see Fig. \ref{fig:Geometrie}). Therefore equation (\ref{eq:1}) describes a global and equation (\ref{eq:2}) a local energy conservation.\\
For the numerical examples in section \ref{sec:Examples}, we have implemented the OMT method developed by Sulman \textit{et al.} \cite{Sul11_1}. It provides a good compromise between an easy implementation and an efficient mapping calculation and is thus sufficient for our test purposes.\\
The result of this design process step is therefore the mapping $\mathbf{u}(x,y)$, but we have to keep in mind, that it is not obvious at this point, whether the OMT mapping is integrable and if it is, for which lens-target distances (see Fig. \ref{fig:Geometrie}) this is the case. This point will be clarified in the next subsection.

\subsection{Freeform surface construction}
\label{sec:Freeform surface construction}

\subsubsection{Ray-mapping condition and surface equation}
\label{sec:Surface equation}

In the following, we want to derive a differential equation for the direct calculation of a freeform surface for a given \textit{general} ray mapping $\mathbf{u}(x,y)$. Our derivations will lead to a condition an \textit{integrable} mapping and its corresponding freeform surface have to fulfill. As we will show, this condition can be fulfilled approximately over a wide range of lens-target distances by the OMT mapping defined in the previous subsection.\\ 
To do so, two basic equations are considered. On the one hand, for an incoming beam described by the ray direction vector field $\mathbf{s}_1$ and the refracted vector field $\mathbf{s}_2$, the law of refraction
\begin{equation}\label{eq:4}
\mathbf{n}=n_1 \mathbf{\hat{s}}_1- n_2 \mathbf{\hat{s}}_2 ,
\end{equation}
with the refractive indices $n_1$ of the lens and $n_2$ of the surrounding medium, has to be fulfilled. On the other hand, we want to ensure the continuity of the surface $z(x,y)$ by the well-known integrability condition \cite{ries02}
\begin{equation}\label{eq:5}
\mathbf{n} \cdot (\nabla \times \mathbf{n})=0.
\end{equation}
Since the collimated beam $\mathbf{s}_1$ as well as $\mathbf{s}_2$ can be expressed in terms of the unknown freeform surface $z(x,y)$ and the given ray mapping $\mathbf{u}(x,y)$ (see Fig. \ref{fig:Geometrie}):
\begin{equation}\label{eq:6}
\mathbf{s}_1 =
\begin{pmatrix}
  0 \\
  0 \\
  z(x,y)
\end{pmatrix}, \ \ \
\mathbf{s}_3 =
\begin{pmatrix}
  u_x (x,y)-x \\
  u_y (x,y)-y \\
  z_T
\end{pmatrix}, \ \ \
\mathbf{s}_2=\mathbf{s}_3-\mathbf{s}_1 , 
\end{equation}
the equations (\ref{eq:4}) and (\ref{eq:5}) represent a differential equation for $z(x,y)$.
Plugging (\ref{eq:4}) into equation (\ref{eq:5}) the integrability condition can be written in the form (see Appendix A)
\begin{equation}\label{eq:7}
\mathbf{s}_2(\nabla \times \mathbf{s}_1)= n_1 \frac{\{\mathbf{s}_2 \times [(\mathbf{s}_2\nabla)\mathbf{s}_2 ]\}_z }{\mathbf{n}\cdot\mathbf{s}_2} +\mathbf{s}_2(\nabla \times \mathbf{s}_3),
\end{equation}
which holds for a \textit{general} ray mapping $\mathbf{u}(x,y)$. Equation (\ref{eq:7}) is organized  in a way that only the left-hand side (LHS) depends on the derivatives of $z(x,y)$. Equation (\ref{eq:7}) takes a more familiar form by inserting the vector fields (\ref{eq:6}), which leads to
\begin{equation}\label{eq:8}
\mathbf{v}\nabla z(x,y)= n_1 \cdot \frac{ \mathbf{v}\cdot \left[\left(\mathbf{v}^{\perp}\cdot \nabla \right)\mathbf{v}^{\perp}\right]}{\mathbf{n}\cdot \mathbf{s}_2}-(z_T- z(x,y) )\nabla \mathbf{v},
\end{equation}
with the velocity field $\mathbf{v}=(\mathbf{u}(x,y)-\mathbf{Id})^{\perp}$, the identity vector $\mathbf{Id}=(x, y)^{T}$ and $\nabla^{\perp}=(-\partial_y, \partial_x)$. This equation is a semilinear two dimensional advection equation, whereby the unknown surface $z(x,y)$ corresponds to conserved transport quantity and the right-hand side (RHS) to a source term.\\
In principle one could try to solve equation (\ref{eq:8}) after applying suitable boundary conditions, but as it will be demonstrated, it is not appropriate for finding a continuous freeform surface. This can be seen easily by considering the condition that the normalized vector field (\ref{eq:4}) has to be equal to the gradient of the surface $z(x,y)$:
\begin{equation}\label{eq:9}
\nabla (z-z(x,y))\stackrel{!}{=} \frac{\mathbf{n}}{(\mathbf{n})_z }
\ \ \ 
\Leftrightarrow
\ \ \
\begin{pmatrix}
  -\partial_x z(x,y) \\
  -\partial_y z(x,y) \\
  1
\end{pmatrix}
\stackrel{!}{=}
\begin{pmatrix}
 -\frac{n_2}{|\mathbf{s}_2|\cdot(\mathbf{n})_z }v_y \\
 \frac{n_2}{|\mathbf{s}_2|\cdot(\mathbf{n})_z }v_x \\
  1
\end{pmatrix}.
\end{equation}
Plugging this relation into the LHS of (\ref{eq:8}), we get $\mathbf{v}\nabla z(x,y)\equiv 0$, which can only be fulfilled if the RHS vanishs:
\begin{equation}\label{eq:10}
n_1 \cdot \frac{ \mathbf{v}\cdot \left[\left(\mathbf{v}^{\perp}\cdot \nabla \right)\mathbf{v}^{\perp}\right]}{\mathbf{n}\cdot \mathbf{s}_2}-(z_T- z(x,y) )\nabla \mathbf{v} \stackrel{!}{\equiv}0.
\end{equation}
The importance of this condition is due to the fact that is has to hold for \textit{every} integrable ray mapping. Since $\mathbf{v}\perp (\mathbf{u}(\mathbf{x,y})-\mathbf{Id})$ and therefore $(\mathbf{u}(x,y)-\mathbf{Id}) || \nabla z(x,y)$ it reflects the nature of the law of refraction, that according to the definitions (\ref{eq:6}) and (\ref{eq:9}) the vectors $\mathbf{s_1}(x,y)$, $\mathbf{s_2}(x,y)$ and $\mathbf{n}(x,y)$ have to lie in the same plane.\\
We now know, that the source term of (\ref{eq:8}) has to vanish, but we are still left with the question, if we can find a way to fullfil condition (\ref{eq:10}). The main task is obviously to find a ray mapping for which relation (\ref{eq:10}) holds, which is nontrivial, since it couples the mapping with the unknown function $z(x,y)$.\\
But if we use the OMT mapping, it follows from its vanishing curl that $\nabla \mathbf{v}=0$. If we use in addition to that, the small-angle approximation
\begin{equation}\label{eq:11}
\mathbf{n}\cdot \mathbf{s}_2 \gg n_1 \cdot \mathbf{v}\cdot \left[\left(\mathbf{v}^{\perp}\cdot \nabla \right)\mathbf{v}^{\perp}\right]
\end{equation}
between the surface normal $\mathbf{n}(x,y)$ and the outgoing ray $\mathbf{s}_2 (x,y)$, we see that the condition (\ref{eq:10}) can be fulfilled approximately by using an OMT map. Because of the fact that $\mathbf{n}\cdot \mathbf{s}_2$  is locally proportional to $z_T -z(x,y)$ in contrast to the RHS, (\ref{eq:11}) can be interpreted as a far field approximation. This also implies, that the integrability of OMT map is only asymptotically exact.\\
Hence for an OMT mapping and the small-angle approximation, we get our final equation
\begin{equation}\label{eq:12}
\mathbf{v}\nabla z(x,y)=\nabla (\mathbf{v} \cdot z(x,y))=0, \ \ \
\mathbf{v}=(\mathbf{u}(x,y)-\mathbf{Id})^{\perp}
\end{equation}
which has to be solved to get the required freeform surface $z(x,y)$.\\

\subsubsection{Boundary conditions}
\label{sec:Boundary conditions}

\begin{figure}[htbp]%
\centering
\includegraphics[width=5.5cm]{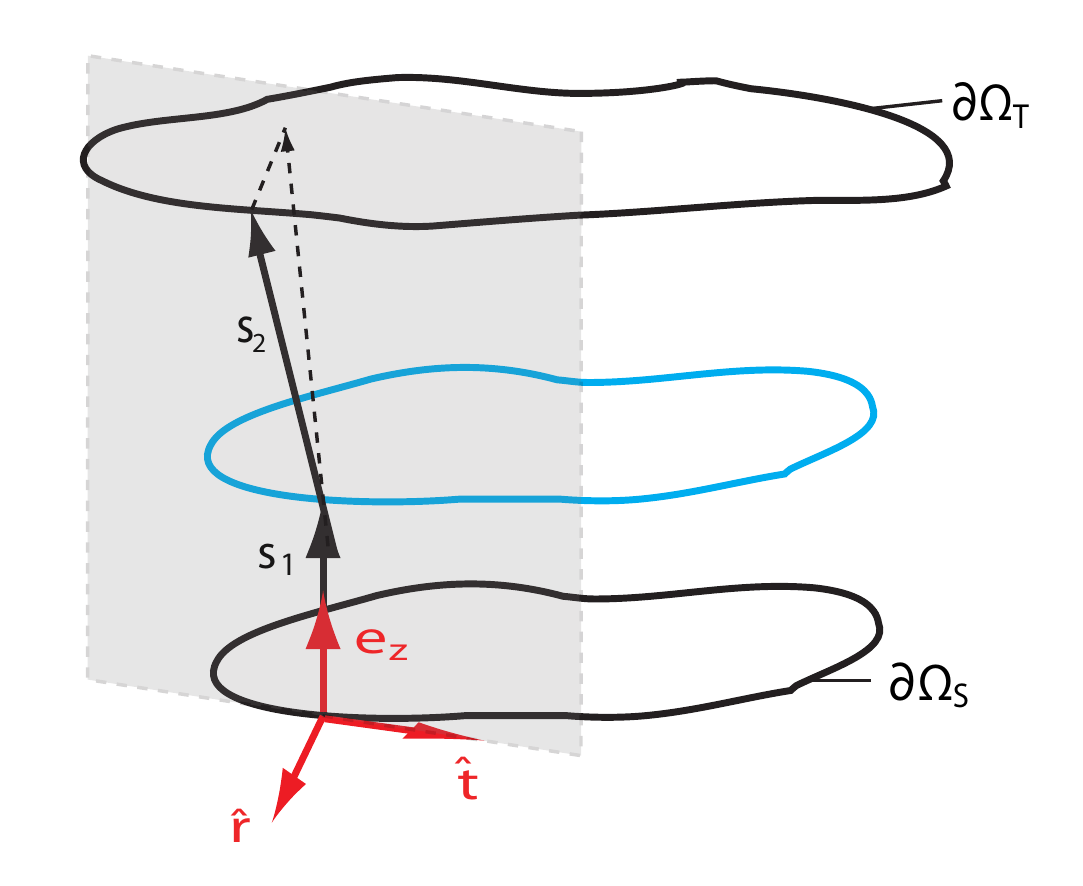}%
\caption{The boundary $\partial \Omega_S$ is parameterized by $s$. At each point of the boundary the local coordinate system is spanned by the tangential vector $\mathbf{\hat{t}}$ and normal vector $\mathbf{\hat{r}}$ to the boundary as well as the unit vector $\mathbf{e}_z$. Since the boundary values $z(s)$ of the freeform surface only determine the tangential deflection of the rays hitting the boundary, the projection of the law of refraction (\ref{eq:4}) on the $\mathbf{\hat{t}}(s)-\mathbf{e}_z -$ plane can be used for the calculation of the boundary values.}
\label{fig:Boundary}
\end{figure}

If we want to solve a linear advection equation like (\ref{eq:12}), we have to know the function $z(x,y)$ on the inflow part of the boundary, where the velocity field points into the integration area $\Omega_S$ \cite{Kuz_1}. Because of energy conservation, this area is defined as $\Omega_S := \overline{ \{(x,y)\in\mathbb{R}_2 \ | \ I_S (x,y)\neq 0 \} }$ and therefore its inflow part by $\partial \Omega_{S-}:=\{ (x,y) \in \partial \Omega_S | \mathbf{v}\cdot \mathbf{\hat{r}}<0 \} $ with the outward boundary normal $\mathbf{\hat{r}}$. Together with this boundary condition equation (\ref{eq:12}) has at most one solution \cite{Kuz_1}.\\
In our case the boundary conditions can be deduced in the following way. First, we have to realize, that for an incoming collimated beam the boundary of the freeform surface can only determine the tangential deflection of a ray refracted at the boundary. The normal deflection is determined by the inner parts of the surface $z(x,y)$. Therefore it makes sense to parameterize the boundary $\partial \Omega_S$ by a parameter $s$ and define the local coordinate system at each point of the boundary by the vectors(see Fig. \ref{fig:Boundary})
\begin{equation}\label{eq:13}
\mathbf{t}=\frac{d}{ds}
\begin{pmatrix}
x(s)  \\
y(s)  \\
0
\end{pmatrix}, \ \ \
\mathbf{r}=\frac{d}{ds}
\begin{pmatrix}
y(s)  \\
-x(s)  \\
0
\end{pmatrix}, \ \ \
\mathbf{e}_z=
\begin{pmatrix}
0  \\
0  \\
1
\end{pmatrix}.
\end{equation}

Since $z(s), s\in \partial \Omega_S$ only determines the tangential deflection, it is sufficient to consider the law of refraction (\ref{eq:4}) in the tangential plane spanned by $\mathbf{\hat{t}}(s)$ and $\mathbf{e}_z $. Hence, we can interpret the boundary value calculation as a two dimensional problem which allows us to derive a differential equation for the boundary values  $z(s)$ by the two dimensional equivalent of equation (\ref{eq:9}) projected on the $\mathbf{\hat{t}}(s)-\mathbf{e}_z -$ plane
\begin{equation}\label{eq:14}
\frac{\mathbf{n}}{(\mathbf{n})_z }\stackrel{!}{=}
\begin{pmatrix}
-\partial_l z(l) \\
1
\end{pmatrix}, \ \ \ \ \
l(s):= \int_0^s \sqrt{\left(\frac{dx}{dt}\right)^2 +\left(\frac{dy}{dt}\right)^2}dt,
\end{equation}
where path length $l(s)$ was introduced for dimensional reasons. From this we get:
\begin{equation}\label{eq:15}
\partial_s z(s)=-\frac{\mathbf{s}_2\cdot \mathbf{t}}{(z_T -z(s))-\frac{n_1}{n_2}\sqrt{(\mathbf{s}_2\cdot \hat{\mathbf{t}})^2 +(z_T -z(s))^2}}
\end{equation}
which reduces in the far field to
\begin{equation}\label{eq:16}
\partial_s z(s)\stackrel{z_T \rightarrow \infty}{\sim} -\frac{\mathbf{s}_2\cdot \mathbf{t}}{z_T\left(1-\frac{n_1}{n_2}\right)}=\frac{v_x \partial_s y - v_y \partial_s x}{z_T\left(1-\frac{n_1}{n_2}\right)} ,
\end{equation}
whereby the position  of the surface in space compared to the target plane is fixed by integration constant. Since (\ref{eq:12}) itself can be interpreted as a far field approximation, as explained above, equation (\ref{eq:16}) seems more suitable for our purposes. It provides us with a simple way of calculating the boundary values. The only degree of freedom left is the integration constant.
Equation (\ref{eq:16}) will build the basis of the numerical algorithm for solving equation (\ref{eq:12}) presented in the next section.\\
At the end of this section, we want to note that (\ref{eq:12}) and (\ref{eq:16}) can be derived analogously for freeform mirrors by replacing (\ref{eq:4}) by the law of reflection and keeping in mind that $z_T < z(x,y)$.

\section{Numerical Algorithm}
\label{sec:Numerical Algorithm}

\begin{figure}[htbp]%
\centering
\includegraphics[width=5.5cm]{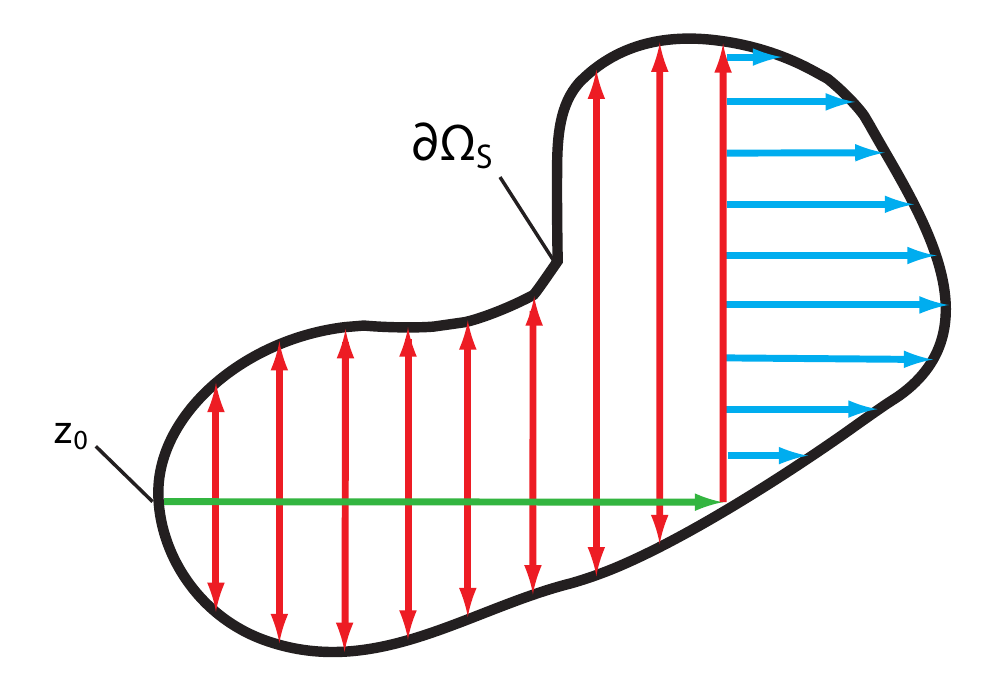}%
\caption{The figure shows one possible way of solving the linear advection equation (\ref{eq:12}) by the simple integration of (\ref{eq:16}) along straight lines. First the value $z_0 =z(x_0 ,y_0 )$ is fixed and used for the integration along the green line. The values of $z(x,y)$ on the green line serve as starting values for the line-by-line integration along the red lines in the orthogonal direction. Then the blue lines are integrated by using the values of $z(x,y)$ on the last red line.}
\label{fig:Integration}
\end{figure}

We could solve equation (\ref{eq:12}) by standard computational fluid dynamic approaches, like finite volume methods, which are appropriate for the numerical treatment of linear advection equations. Based on the nature of equation (\ref{eq:16}), a different approach is proposed in the following.\\
Considering (\ref{eq:16}), we recognize that the boundary values are calculated by the velocity field itself. This is in contrast to the usual fluid dynamical framework and allows us to separate $\Omega_S$ into an arbitrary number of subareas $\Omega_{S,i}$ for each of which we can calculate the boundary values by (\ref{eq:16}). Therefore, the freeform surface can be calculated on each subarea $\Omega_{S,i}$ and the solution on $\Omega_{S}$ by their unification.
This implies that the freeform surface can be constructed by an integration of the OMT map along arbitrary paths on $\Omega_{S}$, which characterizes the integrability of ray mappings \cite{Fou10_1, Ma15_1}. Thus according to (\ref{eq:12}) and (\ref{eq:16}) the OMT map is approximately integrable as long as (\ref{eq:11}) holds.\\
Hence, the most convenient way to get the solution of (\ref{eq:12}) seems to be a line-by-line integration of (\ref{eq:16}), which along the $x$- and $y$-direction is equivalent to equation (\ref{eq:9}) in a far field approximation. One possible way of integrating (\ref{eq:16}) is shown as an example in Fig. \ref{fig:Integration}. Thus, only the integration constant of one integral has to be fixed from which the others follow automatically.\\ 
The proposed approach has the useful feature that we do not need to parameterize the boundary $\partial \Omega_{S}$, which allows the calculation of freeform surfaces with complex boundary shapes.
The efficiency of the line-by-line integration approach is shown in the next chapter for two challenging design examples.

\section{Examples}
\label{sec:Examples}

\begin{figure*}
\centering
\includegraphics[width=\textwidth]{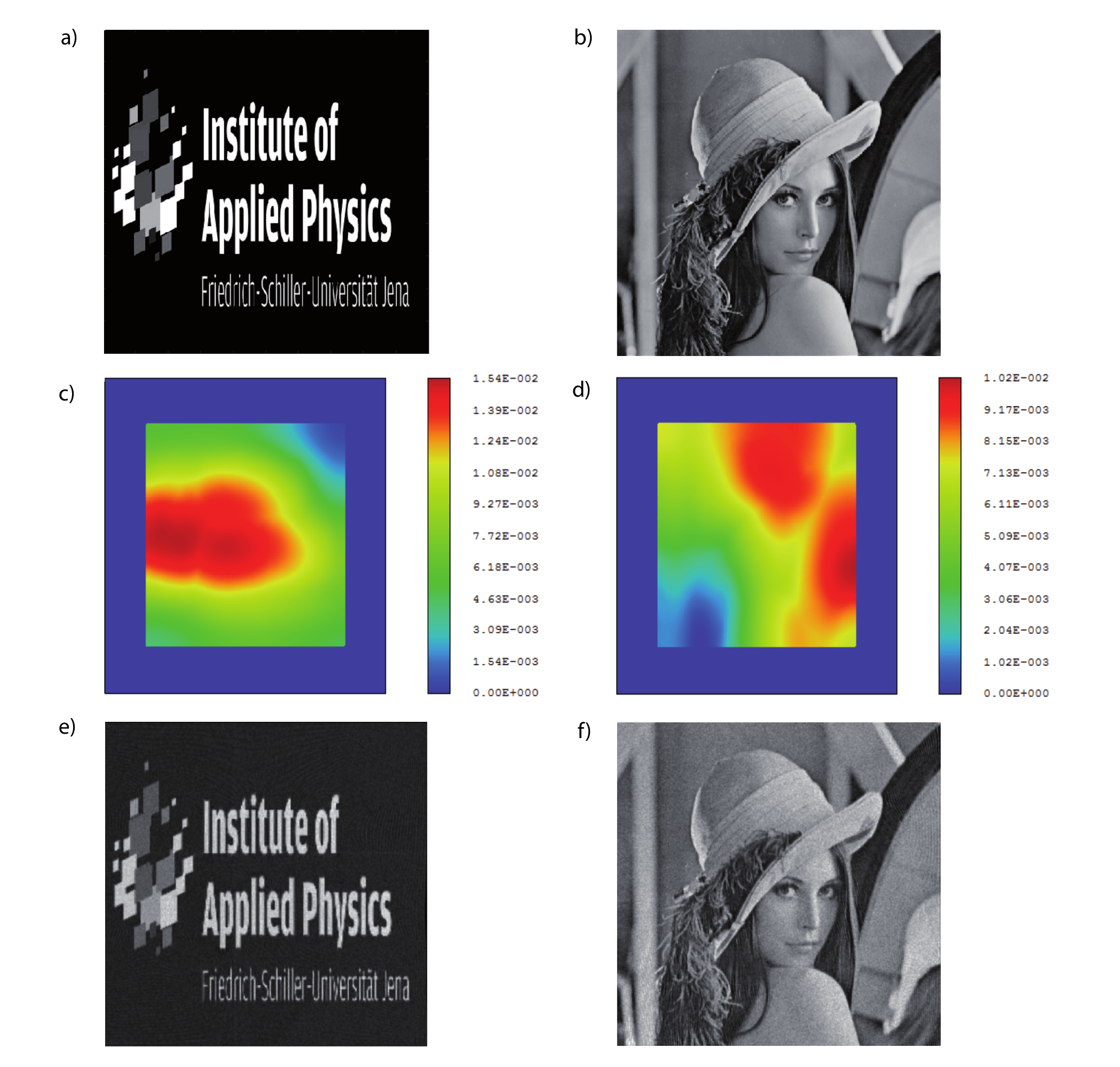}\\ 
\caption{a), b): given output intensities. In both cases, the incoming collimated beam was chosen to have a uniform intensity.
c), d): false color maps of the calculated freeform lenses with side length of one. The integration constant of equation (\ref{eq:16}) at the upper left side was  $z_0 =1$ and the source-target distance $z_T =5$. The imported numerical data was interpolated automatically by ZEMAX.
e), f): intensity pattern from a ZEMAX raytracing with $4\cdot 10^{7}$ rays.}
\label{fig:IAP}
\end{figure*}

To show the efficiency of the algorithm, we want to apply it to two design examples. In the first one, we will calculate a freeform lens that maps a collimated beam of uniform intensity on the logo of the Institute of Applied Physics (IAP) in Jena with a resolution of 500 x 500 pixels (see Fig. \ref{fig:IAP}). It shows strong intensity gradients between the letters and the background. To omit a division by zero within the implemented OMT algorithm \cite{Sul11_1} we have to use a background intensity $I>0$ for the input and output intensities. For an appropriate speed of convergence the background intensity is set to $20$ per cent of the maximum intensity. The second example, which shows smooth intensity variations and a lot of details, is the well-known picture of Lena with a resolution of 500 x 500 pixels (see Fig. \ref{fig:IAP}).\\
Since the freeforms were calculated by integrating equation (\ref{eq:16}), the specific characteristics of both pictures do not have any influence on the speed of the lens construction step explained in section \ref{sec:Numerical Algorithm}, but they increase the mapping calculation time.\\
The pictures both have a square format. Hence, it is convenient to integrate first along the upper side of the square region and then line-by-line in the orthogonal direction with the starting values given by the first integration. Therefore we have to solve 501 integrals, which took in both cases less than one second in MATLAB on an Intel Core i3 at 2.4Ghz with 16GB RAM. This time has to be added up with the mapping calculation time, which strongly depends on the implemented method and the specific features of the picture.\\
The integration constant on the upper left side of the integration area was chosen to be $z_0=1$. Values of $n_1=1.5$ and $n_2 =1$ were used for the refractive indices, and the source-target distance was chosen to be $z_T =5$. The input and output beam as well as the freeform lens had side length of one. Since every spatial value is normalized the results are scalable.\\
At this point we want to note that according to (\ref{eq:16}) the validity of the approximation (\ref{eq:11}) can simply be checked by scaling the numerical results with $1/ z_T$, which of course has to be done for each intensity and configuration individualy. For our examples the quality of the illumination patterns produced by the raytracing did not change significantly even for distances between the lens and the target plane smaller than the side length of the lens. \\
In both cases the calculated freeform lens was imported as a grid sag surface into ZEMAX to verify our results by a raytracing simulation. The imported lens data are interpolated by ZEMAX automatically. The results can be seen in Fig. \ref{fig:IAP}.

\section{Conclusion}
\label{sec:Conclusion}

We presented an efficient numerical method for the single freeform surface design for the shaping of collimated beams. It is based on the derivation of the condition (\ref{eq:10}) for integrable mappings by combining the law of reflaction/refraction and the well-known integrability condition in a suitable way. We showed that the condition can be fulfilled in the small-angle approximation (\ref{eq:11}) by using a ray mapping calculated by optimal mass transport. Equation (\ref{eq:11}) therefore represents a quantitative estimate for the applicability of the OMT map. This serves as a theoretical basis for the decoupling of the design process into two separate steps: the calculation of the OMT mapping and the construction of the freeform surface with a linear advection equation. On the basis of the finding of appropriate boundary conditions for the advection equation, we presented a simple numerical algorithm for the surface construction by solving the standard integrals (\ref{eq:16}), which differs from the previously published OMT freeform design methods.\\
Besides its simplicity, accuracy and quickness, a useful feature of the construction technique is the independence of the freeform boundary shape (see Fig. \ref{fig:Integration}). By using a proper method for the calculation of the OMT mapping, this allows for example the calculation of a freeform for disconnected intensities or source and target intensties with two concave boundaries, which is in general a nontrivial problem, but important for applications.\\
The results imply the approximate integrability of the OMT map over a wide range of freeform-target plane distances and gives the OMT freeform design methods a theoretical basis for collimated input beams. Especially the condition (\ref{eq:10}) is interesting, also for other ray mapping techniques than OMT methods, since it has to hold for every integrable mapping. Hence, it opens up new possibilities for nearfield calculations as well as generalizations to it for point sources and double freeforms, which are currently under work.

\newpage

\appendix

\section*{Appendix A}
\label{sec:Appendix A}
\setcounter{equation}{0}
\renewcommand{\theequation}{A{\arabic{equation}}}

We want to derive equation (\ref{eq:8}) from the law of refraction (\ref{eq:4}) and the integrability condition (\ref{eq:5}). Since the curl of the incident field $\mathbf{\hat{s}}_1$ vanishs, plugging (\ref{eq:4}) into (\ref{eq:5}) gives
\begin{equation}\label{eq:A.1}
\mathbf{n} \cdot \left(\nabla \times \frac{\mathbf{s}_2}{|\mathbf{s}_2|}\right)=\mathbf{n} \cdot \left(\frac{1}{|\mathbf{s}_2|}\nabla \times \mathbf{s}_2 -\mathbf{s}_2 \times \nabla \frac{1}{|\mathbf{s}_2|}\right)=0
\end{equation}
With
\begin{equation}\label{eq:A.2}
\nabla \frac{1}{|\mathbf{s}_2|}= -\frac{1}{2|\mathbf{s}_2|^{3}}\nabla(\mathbf{s}_2 \cdot \mathbf{s}_2)=-\frac{1}{|\mathbf{s}_2|^3}\left[\mathbf{s}_2 \times (\nabla \times \mathbf{s}_2 ) +(\mathbf{s}_2\nabla)\mathbf{s}_2 \right]\\		\nonumber
\end{equation}
it follows
\begin{equation}\label{eq:A.21}
\mathbf{s}_2 \times \nabla \frac{1}{|\mathbf{s}_2|} = -\frac{1}{|\mathbf{s}_2|^3}\{\underbrace{\mathbf{s}_2 \times [\mathbf{s}_2 \times (\nabla \times \mathbf{s}_2)]}_{=[\mathbf{s}_2(\nabla \times \mathbf{s}_2)]\mathbf{s}_2-|\mathbf{s}_2 |^2 (\nabla \times \mathbf{s}_2)} +\mathbf{s}_2 \times [(\mathbf{s}_2\nabla)\mathbf{s}_2 ] \}
\end{equation}
and we can write (\ref{eq:A.1}) as
\begin{equation}\label{eq:A.3}
 \mathbf{n} \cdot \left\{[\mathbf{s}_2(\nabla \times \mathbf{s}_2)]\mathbf{s}_2 + \mathbf{s}_2 \times [(\mathbf{s}_2\nabla)\mathbf{s}_2 ]\right\} =0. 
 \end{equation}
Inserting (\ref{eq:4}) and using $\mathbf{\hat{s}}_1 =(0,0,1)$ and $\mathbf{\hat{s}}_2\cdot (\mathbf{s}_2 \times ...)=0$ leads to
\begin{equation}\label{eq:A.4}
 (\mathbf{n}\cdot \mathbf{s}_2) \cdot [\mathbf{s}_2(\nabla \times \mathbf{s}_2)] + n_1 \{\mathbf{s}_2 \times [(\mathbf{s}_2\nabla)\mathbf{s}_2 ]\}_z =0
 \end{equation}
and with $\mathbf{s}_2=\mathbf{s}_3 -\mathbf{s}_1$ it follows
\begin{equation}\label{eq:A.5}
\mathbf{s}_2(\nabla \times \mathbf{s}_1)= n_1 \frac{\{\mathbf{s}_2 \times [(\mathbf{s}_2\nabla)\mathbf{s}_2 ]\}_z }{\mathbf{n}\cdot\mathbf{s}_2} +\mathbf{s}_2(\nabla \times \mathbf{s}_3). 
\end{equation}
Using the definition $\mathbf{v}=(-(\mathbf{s}_2)_y ,(\mathbf{s}_2)_x )$ and (\ref{eq:6}) the terms in equation (\ref{eq:A.5}) can be written as
\begin{equation}\label{eq:A.6}
\mathbf{s}_2(\nabla \times \mathbf{s}_1)= 
\begin{pmatrix}
-(\mathbf{s}_2)_y  \\
(\mathbf{s}_2)_x 
\end{pmatrix}\cdot
\begin{pmatrix}
\partial_x z(x,y)  \\
\partial_y z(x,y)
\end{pmatrix}
=\mathbf{v} \nabla z(x,y)
\end{equation}
and
\begin{equation}\label{eq:A.7}
\begin{split}
\{\mathbf{s}_2 \times [(\mathbf{s}_2 \cdot \nabla)\mathbf{s}_2]\}_z 
& = (\mathbf{s}_2)_x \cdot [(\mathbf{s}_2 \cdot \nabla)\mathbf{s}_2]_y - (\mathbf{s}_2)_y  \cdot [(\mathbf{s}_2 \cdot \nabla)\mathbf{s}_2]_x  \\
& =
\begin{pmatrix}
-(\mathbf{s}_2)_y  \\
(\mathbf{s}_2)_x 
\end{pmatrix}\cdot
\left[(\mathbf{s}_2 \cdot \nabla) 
\begin{pmatrix}
(\mathbf{s}_2)_x  \\
(\mathbf{s}_2)_y 
\end{pmatrix}
\right] \\
& = \mathbf{v}\cdot \left[\left(\mathbf{v}^{\perp}\cdot \nabla \right)\mathbf{v}^{\perp}\right]
\end{split}
\end{equation}
and
\begin{equation}\label{eq:A.8}
\begin{split}
\mathbf{s}_2(\nabla \times \mathbf{s}_3)
& =(z_T-z(x,y))\cdot [\partial_x (\mathbf{s}_2)_y - \partial_y (\mathbf{s}_2)_x] \\
& =-(z_T-z(x,y))\nabla \mathbf{v}.
\end{split}
\end{equation}

\section*{Acknowledgments}

The authors thank M. Esslinger, and M. Tessmer for valuable discussions, R. Hambach and  S. Schmidt for valuable discussions and comments on the manuscript, C. Liu and D. Lokanathan for the help with the ZEMAX implementation and D. Musick for the spelling and grammar check. We also
acknowledge the Federal Ministry of Education and Research Germany for financial support through the project KoSimO (FKZ:031PT609X).


\begin{thebibliography}{1}


\bibitem{ries02}
H. Ries and J. Muschaweck, ``Tailored freeform optical surfaces,''
J. Opt. Soc. Am. A {\bf 19}(3), 590--595 (2002).
 
\bibitem{Wu13_1}
R. Wu, L. Xu, P. Liu, Y. Zhang, Z. Zheng, H. Li, and X. Liu, ``Freeform illuminaton design: a nonlinear boundary problem for the elliptic Monge-Ampère equation,''
Opt. Lett. {\bf 38}(2), 229--231 (2013).
  
\bibitem{Wu13_2}
R.~Wu, K.~Li, P.~Liu, Z.~Zheng, H.~Li, and X.~Liu, ``Conceptual design of dedicated road lighting for city park and housing estate,''
 Appl. Opt. {\bf 52}(21), 5272--5278 (2013).
 
\bibitem{Wu13_3}
R.~Wu, P.~Liu, Y.~Zhang, Z.~Zheng, H.~Li, and X.~Liu, ``A mathematical model of the single freeform surface design for collimated beam shaping,''
 Opt. Express {\bf 21}(18), 20974--20989 (2013). 

\bibitem{Wu14_1}
Y.~Zhang, R.~Wu, P.~Liu, Z.~Zheng, H.~Li, and X.~Liu, ``Double freeform surfaces design for laser beam shaping with Monge-Ampère equation method,''
Opt. Comm. {\bf 331}, 297--305 (2014). 

\bibitem{Wu14_2}
R.~Wu, Y.~Zhang, M.M.~Sulman, Z.~Zheng, P.~Benítez, and J.C.~Miñano,  ``Initial design with L2 Monge-Kantorovich theory for the Monge–Ampère equation method in freeform surface illumination design,'' Opt. Express {\bf 22}(13), 16161--16177 (2014) 

\bibitem{Prins14_1}
C.R.~Prins. J.H.M.~ten Thije Boonkkamp, J.~Van Roosmalen, W.L. IJzerman, and T.W.~Tukker, ``A Monge-Ampère-Solver for free-form reflector design,''
SIAM J. Sci. Comput. {\bf 36}(3), B640--B660 (2014). 

\bibitem{Brix15_1}
K.~Brix, Y.~Hafizogullari, and A.~Platen, ``Designing illumination lenses and mirrors by the numerical solution of Monge–Ampère equations,'' 
J. Opt. Soc. Am. A {\bf 32}(11), 2227-2236 (2015)

\bibitem{Oliker02_1}
V.I.~Oliker, ``Mathematical aspects of design of beam shaping surfaces in geometrical optics,''
in \textit{Trends in Nonlinear Analysis}, M. Kirkilionis, S. Kromker, R. Rannacher, and F. Tomi, eds. (Springer-Verlag, 2003), pp. 193-222.

\bibitem{Fou10_1}
F.R.~Fournier, W.J.~Cassarly, and J.P.~Rolland, ``Fast freeform reflector generation using source-target maps,''
Opt. Express {\bf 18}(5), 5295--5304 (2010).

\bibitem{Can13_1}
C.~Canavesi, W.J.~Cassarly, and J.P.~Rolland, ``Target flux estimation by calculating intersections between neighboring conic reflector patches,''
Opt. Lett. {\bf 38}(23), 5012-5015 (2013).

\bibitem{Ma15_1}
D.~Ma, Z.~Feng, and R.~Liang, ``Tailoring freeform illumination optics in a double-pole coordinate system,''
Appl. Opt. {\bf 54}(9), 2395-2399 (2015) 

\bibitem{Mic11_1}
D.~Michaelis, D.~Schreiber, and A.~Bräuer, ``Cartesian oval representation of freeform optics in illumination systems,''
Opt. Lett. {\bf 36}(6), 918--920 (2011).

\bibitem{Rub07_1}
J. Rubinstein, and G. Wolansky, ``Intensity control with a free-form lens,'' J. Opt. Soc. Am. A {\bf 24}(2), 463-469 (2007) .

\bibitem{Oli13_1}
V.~Oliker, J.~Rubinstein, and G.~Wolansky, ``Ray mapping and illumination control,'' J. Photon. Energy  {\bf 3}(1), 035599 (2013).

\bibitem{Oli14_1}
V.~Oliker, ``Differential equations for design of a freeform single lens with prescribed irradiance properties,'' Opt. Eng. {\bf 53}(3), 031302 (2014).

\bibitem{Oli14_2}
V.~Oliker, and  B.~Cherkasskiy, ``Controlling light with freeform optics: recent progress in computational methods for optical design of freeform  lenses with prescribed irradiance properties,'' Proc. SPIE {\bf 9191}, 919105--919105-7 (2014).


\bibitem{Bau12_1}
A.~B\"auerle, A.~Bruneton, P.~Loosen, and R.~Wester, ``Algorithm for irradiance tailoring using multiple freeform optical surfaces,''
Opt. Express {\bf 20}(13), 14477--14485 (2012). 

\bibitem{Brun13_1}
A.~Bruneton, A.~B\"auerle, P.~Loosen, and R.~Wester, ``High resolution irradiance tailoring using multiple freeform surfaces,''
Opt. Express {\bf 21}(9), 10563--10571 (2013). 

\bibitem{Feng13_1}
Z.~Feng, L.~Huang, G.~Jin, and M.~Gong, ``Beam shaping system design using double freeform optical surfaces,''
Opt. Express {\bf 21}(12), 14728--14735 (2013). 

\bibitem{Feng13_2}
Z.~Feng, L.~Huang, G.~Jin, and M.~Gong, ``Designing double freeform optical surfaces for controlling both irradiance and wavefront,''
Opt. Express {\bf 21}(23), 28693--28701 (2013). 

\bibitem{Oli11_1}
V.~Oliker, ``Designing freeform lenses for intensity and phase control of coherent light with help from geometry and mass transport,'' Arch. Ration. Mech. Anal. {\bf 201}(3), 1013-1045 (2011).

\bibitem{Bre02_1}
J.-D.~Benamou, Y.~Brenier, and K.~Guittet, ``The Monge-Kantorovich mass transfer and its Compuational Fluid Mechanics formulation,''
Int. J. Numer. Meth. Fluids {\bf 40}(1-2), 21--30 (2002).

\bibitem{Hak04_1}
S.~Haker, L.~Zhu, A.~Tannenbaum, and S.~Angenent, ``Optimal mass transport for registration and warping,''
Int. J. Comp. Vis.  {\bf 60}(3), 225-240 (2004).

\bibitem{Sul11_1}
M.M.~Sulman, J.F.~Williams, and R.D.~Russel, ``An efficient approach for the numerical solution of Monge-Ampère equation,''
Appl. Numer. Math. {\bf 61}(3), 298-307 (2011).

\bibitem{Kuz_1}
D.~Kuzmin,
\textit{A Guide to Numerical Methods for Transport Equations} (University Erlangen-Nuremberg, 2010).

\end{thebibliography}
\end{document}